\documentstyle[12pt,preprint,aps,floats]{revtex}
\def\NPB#1#2#3{Nucl. Phys. {\bf B#1}, #3 (19#2)}

\def\PLB#1#2#3{Phys. Lett. {\bf B#1}, #3 (19#2)}

\def\PRD#1#2#3{Phys. Rev. {\bf D#1}, #3 (19#2)}
\def\PRL#1#2#3{Phys. Rev. Lett. {\bf#1}, #3 (19#2)}

\input epsf.tex
\def\DESepsf(#1 width #2){\epsfxsize=#2 \epsfbox{#1} \vspace*{0.07in}}
\newcommand{\postscript}[2]{\setlength{\epsfxsize}{#2\hsize}
   \centerline{\epsfbox{#1}}}

\newcommand{\met}{\not{\hbox{\kern-4pt $E_T$}}}
\newcommand{\wino}{\tilde{W}}
\newcommand{\chargino}{\tilde{\chi}}
\newcommand{\neutralino}{\tilde{\chi}^0}

\newcommand{\ifb}{ \text{ fb}^{-1}}
\newcommand{\tev}{\text{ TeV}}
\newcommand{\gev}{\text{ GeV}}
\newcommand{\mev}{\text{ MeV}}
\newcommand{\Dzero}{\text{D\O}}
\tighten

\begin{document}

\preprint{
\noindent
\begin{minipage}[t]{3in}
\begin{flushleft}
April 1999 \\
\end{flushleft}
\end{minipage}
\hfill
\begin{minipage}[t]{3in}
\begin{flushright}
IASSNS--HEP--99--19\\
MIT--CTP--2830\\
PUPT--1857\\
hep-ph/9904250\\
\vspace*{.7in}
\end{flushright}
\end{minipage}
}

\title{Discovering Supersymmetry at the Tevatron in Wino LSP
Scenarios}

\author{Jonathan L. Feng,$^a$ Takeo Moroi,$^a$ Lisa Randall,$^{bc}$
Matthew Strassler,$^a$ and Shufang Su$^c$
\vspace*{.2in}
}
\address{${}^{a}$School of Natural Sciences,
Institute for Advanced Study, Princeton, NJ 08540 USA}
\address{${}^{b}$Joseph Henry Laboratories, Princeton University,
Princeton, NJ 08543 USA}
\address{${}^{c}$Center for Theoretical Physics, Massachusetts
Institute of Technology, Cambridge, MA 02139 USA
\vspace*{.2in}
}

\maketitle

\begin{abstract}
In supersymmetric models, Winos, partners of the SU(2) gauge bosons,
may be the lightest supersymmetric particles (LSPs). For generic
parameters, charged and neutral Winos are highly degenerate.  Charged
Winos travel macroscopic distances, but can decay to neutral Winos and
extremely soft leptons or pions before reaching the muon chambers,
thereby circumventing conventional trigger requirements based on
energetic decay products or muon chamber hits.  However, these
charginos are detectable, and can be triggered on when produced in
association with jets.  In addition, we propose a new trigger for
events with a high $p_T$ track and low hadronic activity.  For
Tevatron Run II with luminosity 2 fb$^{-1}$, the proposed searches can
discover Winos with masses up to 300 GeV and explore a substantial
portion of the parameter space in sequestered sector models \cite{lr}.

\vspace*{-.4in}
\end{abstract}

\pacs{PACS numbers: 14.80.Ly, 11.30.Er, 12.60.Jv, 11.30.Pb}


The discovery of supersymmetry (SUSY) is much anticipated at high
energy colliders.  If SUSY is to retain its motivation of stabilizing
the electroweak scale against large radiative corrections, at least
some supersymmetric particles must have masses of order the
electroweak scale.  In the most widely studied models, the lightest
supersymmetric particle (LSP) is assumed to be stable and the partner
of the U(1)$_Y$ gauge boson.  SUSY signals are then characterized by
missing transverse energy ($\met$) and are unlikely to escape
detection when the Large Hadron Collider (LHC) at CERN begins
operation around 2005 with center of mass energy $\sqrt{s} = 14\tev$.

Recently, however, it has been realized that many other SUSY
signatures are possible.  While these signatures vary widely, a number
of them are in fact even more striking than the classic $\met$
signature and give new life to the hope that the discovery of SUSY
need not wait for the LHC~\cite{stauth,stauexp,Chen}.  In this letter,
we study scenarios in which the LSP, while still the lightest
neutralino, is not the U(1)$_Y$ gaugino, but the neutral SU(2)
gaugino, the Wino $\wino^0$.  We will see that this simple
modification leads to drastic differences in phenomenology. These were
argued to make detection difficult, based on conventional triggers, in
\cite{Chen,giudice}, but were argued to provide a novel identifiable
signal in Ref.~\cite{lr}. In this paper, we elaborate on this
observation.  As in more conventional scenarios, the neutral LSP
interacts very weakly and escapes detection.  The new element is that
the next-to-lightest superpartner, the charged Wino $\wino^{\pm}$, is
generically extremely degenerate with the LSP and decays after
centimeters or meters to an LSP and an extremely soft lepton or pion.
Such charged Winos are therefore missed by conventional triggers and
avoid detection in traditional searches.  However, if care is taken to
preserve such events at the trigger level, we will see that large and
spectacular signals may appear at the upcoming run of the Fermilab
Tevatron with $\sqrt{s} = 2 \tev$.

At tree-level, the masses of the charginos and neutralinos depend on
the U(1)$_Y$ gaugino mass $M_1$, the SU(2) gaugino mass $M_2$, the
Higgsino mass $\mu$, and $\tan\beta$, the ratio of Higgs vacuum
expectation values. Without loss of generality, we choose $M_2$ real
and positive.  Phases in the parameters $M_1$ and $\mu$ are then
physical.  We will consider the case $M_2 < |M_1|, |\mu|$, so that the
lightest charginos and neutralinos, $\chargino_1^\pm$ and
$\neutralino_1$, are Wino-like with masses $\sim M_2$.  We assume that
all other superparticles are (much) heavier than the Winos.  With this
assumption we may neglect corrections to charged Wino decay from
virtual supersymmetric particles.

We will consider two Wino LSP scenarios.  In the first, we consider
the well-motivated sequestered sector models~\cite{lr}, in which there
is an anomaly-mediated spectrum of gauginos and a consistent scenario
involving light scalars.  In these models, the gaugino mass parameters
are given by \cite{lr,giudice}
\begin{equation} M_i = -b_i g_i^2 M_{\text{SUSY}} \ ,
\end{equation}
where $M_{\text{SUSY}}$ determines the overall SUSY-breaking scale,
$i=1,2,3$ identifies the gauge group, $g_i$ are gauge coupling
constants, and $b_i$ are the 1-loop $\beta$-function coefficients of
the (full supersymmetric) theory.  Substituting the weak scale values
of $g_i$, we find $M_1 : M_2 : M_3 = 3.3 : 1 : -10$.  It should be
borne in mind that sequestered sector models predict a large hierarchy
between Wino and squark masses. Naturalness bounds therefore suggest
$M_2 \alt 200-300$ GeV, and we will see that a large portion of the
parameter space in these scenarios may be explored at the Tevatron.

More generally, the Wino LSP scenario may be realized for a large
region of SUSY parameter space if the assumption of gaugino mass
unification is relaxed~\cite{Chen,Cheng}.  We will therefore also
consider an alternative set of parameters with $M_1 = -1.5 M_2$.  As
will be seen, this choice leads to significant differences from the
anomaly-mediated case, and so serves as an illustrative
alternative. Since these parameters are not motivated by any model,
the Wino mass $M_2$ is less constrained in this case.

The SUSY signal depends strongly on $\Delta M \equiv
m_{\chargino^{\pm}_1} - m_{\neutralino_1}$.  At tree level, the
chargino mass matrix is

\begin{equation}
\label{chamass}
\left( \begin{array}{cc}
 M_2                    &\sqrt{2} \, m_W s_{\beta} \\
\sqrt{2} \, m_W c_{\beta}   &\mu   \end{array} \right)
\end{equation}
in the basis $(-i\tilde{W}^{\pm}, \tilde{H}^{\pm})$, and the
neutralino mass matrix is

\begin{equation}
\label{neumass}
\left( \!\!\! \begin{array}{cccc}
M_1        &0       & -m_Z c_{\beta}\, s_W & m_Z s_{\beta}\, s_W \\
0          &M_2     & m_Z c_{\beta}\, c_W &-m_Z s_{\beta}\, c_W \\
-m_Z c_{\beta}\, s_W  & m_Z c_{\beta}\, c_W &0       &-\mu      \\
 m_Z s_{\beta}\, s_W  & -m_Z s_{\beta}\, c_W &-\mu    &0 \end{array}
\!\!\! \right)
\end{equation}
in the basis $(-i\tilde{B},-i\tilde{W}^3, \tilde{H}^0_1,
\tilde{H}^0_2)$.  Here $s_W\equiv\sin \theta_W$, $c_W\equiv\cos
\theta_W$, $s_{\beta} = \sin\beta$, and $c_{\beta} = \cos\beta$.

The mass matrices may be diagonalized exactly, but it is enlightening
to consider a perturbation series in $1/\mu$ for large $|\mu|$.  The
lightest chargino and neutralino are degenerate at zeroth order with
$m_{\chargino^{\pm}_1}^{(0)} = m_{\neutralino_1}^{(0)} = M_2$.  At the
next order in $1/\mu$, they receive corrections from mixing with the
Higgsinos.  However, both masses are corrected by
$m_{\chargino^{\pm}_1}^{(1)} = m_{\neutralino_1}^{(1)} = - m_W^2 \sin
2\beta / \mu$, so the degeneracy remains.  It is only at the next
order, where the neutralino mass receives contributions from U(1)$_Y$
gaugino mixing which have no counterpart in the chargino sector, that
the degeneracy is broken:

\begin{equation}
\Delta M_{\rm tree}
\approx m_{\chargino^{\pm}_1}^{(2)} - m_{\neutralino_1}^{(2)}
= \frac{m_W^4 \tan^2 \theta_W}{(M_1-M_2) \mu^2} \sin^2 2\beta \ .
\label{tree}
\end{equation}
Note that for large $\tan\beta$, even this contribution is suppressed.
In fact, for $\tan\beta \to \infty$, $\Delta M_{\rm tree} \propto
1/\mu^4$.  (A $1/\mu^3$ contribution vanishes because, in this limit,
the bilinear Higgs scalar coupling $B$ vanishes, and so an exact
Peccei-Quinn symmetry relates $\mu \leftrightarrow -\mu$.)  For all of
these reasons, the mass splitting is highly suppressed, even for
moderate values of $|\mu|$.

Given the large suppression of $\Delta M_{\rm tree}$, 1-loop
contributions may be important.  The leading contribution to the mass
splitting from loop effects is from custodial SU(2)-breaking in the
gauge boson sector.  (Loop contributions from sleptons and squarks are
insignificant for heavy top and bottom squarks~\cite{Cheng}.)  The
loop contribution is positive, and, in the pure Wino limit, it has the
simple form~\cite{Cheng}, letting $r_i = m_i/M_2$,

\begin{equation}
\Delta M_{\rm 1-loop} = \frac{\alpha_2 M_2}{4\pi} \left[
f(r_W) - c_W^2 f(r_Z) - s_W^2 f(r_\gamma) \right] ,
\end{equation}
where $f(a)=\int_0^1 dx (2 + 2x) \log [x^2 + (1-x) a^2]$.

\begin{figure}[tb]
\postscript{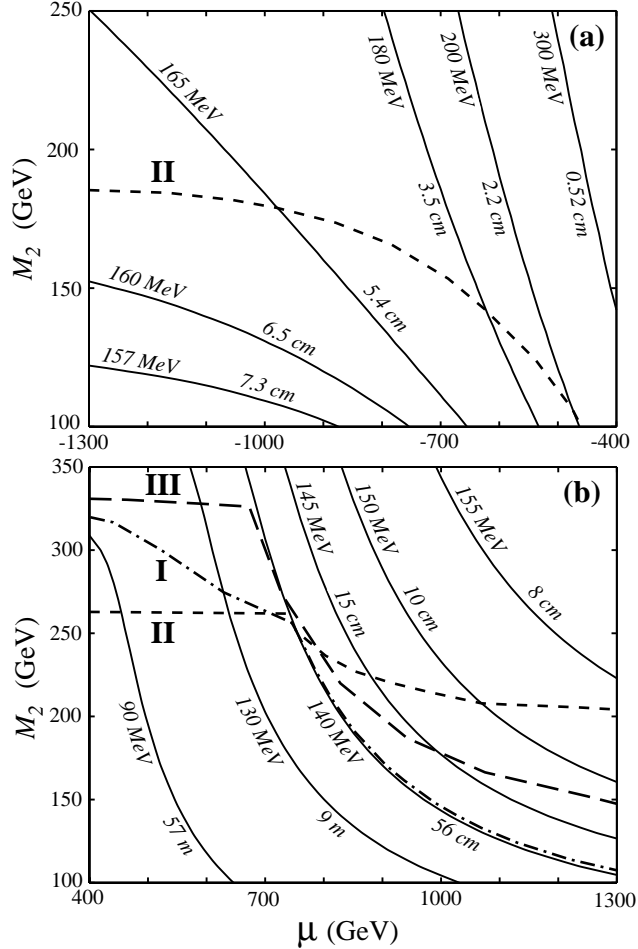}{0.51}
\caption{The mass splitting $\Delta M \equiv m_{\chargino^{\pm}_1} -
m_{\neutralino_1}$ and decay lengths $c\tau$ in the $(\mu, M_2)$
plane. (a) The anomaly-mediated relation $M_1 \approx 3.3 M_2$ is
assumed, and $\tan\beta = 10$. Similar results are obtained for $\mu
>0$. The discovery region for trigger II is shown. (See text.)  (b)
The same for a more general Wino LSP model, with $\tan\beta=3$ and
$M_1=-1.5 M_2$, along with the discovery reach for triggers I --
III. (See text.)}
\label{fig:deltam}
\end{figure}

In Fig.~\ref{fig:deltam}a, we plot the total mass splitting $\Delta M$
for the anomaly-mediated value of $M_1/M_2$ and a moderate value of
$\tan\beta$, where the tree-level mass matrices have been corrected by
1-loop gauge boson contributions including chargino and neutralino
mixing~\cite{Pierce} and have been diagonalized numerically.  We show
the region (for $\mu<0$) of parameter space which is consistent with
naturalness constraints~\cite{fm}.  Typical mass splittings are of
order 150 MeV to 1 GeV.  In Fig.~\ref{fig:deltam}b we do the same for
a model with  $M_1=-1.5M_2$, in which $\Delta M$ may be even
smaller.  Note that the near-degeneracy of the Wino-like chargino and
neutralino is generic.  Generally, this degeneracy is not of great
phenomenological importance, as the Wino-like chargino and neutralino
both decay quickly to other particles.  However, when one of them is the LSP,
the other must decay into it, and the near-degeneracy results in
macroscopic decay lengths with important implications.

For mass splittings in the range of a few hundred MeV, the dominant
chargino decays are the three-body decays $\chargino^+_1 \to
\neutralino_1 (e^+ \nu_e, \mu^+ \nu_\mu)$, and the two-body decay
$\chargino^+_1 \to \neutralino_1 \pi^+$.  For $\Delta M \alt
m_{\pi^{\pm}} \simeq 140\mev$, the decay rate is dominated by the
electron mode, with $\Gamma( \chargino^+_1 \to \neutralino_1 e^+ \nu )
\approx \frac{G_F^2}{(2\pi)^3} \frac{16}{15} (\Delta M)^5$,
corresponding to a decay length of $c\tau|_{e \text{ mode}} = 34
\text{ meters} \times \left(100 \mev / \Delta M \right)^5$ \cite{lr}.
However, once the pion mode becomes available, it quickly
dominates~\cite{Chen,tw}, and $c\tau$ becomes of order 10 cm or less.
In Fig.~\ref{fig:deltam}, the contours are labeled also with decay
lengths $c\tau$, where all final states are included.  We find
macroscopic decay lengths on the order of centimeters to meters in
much of the parameter space.

Amazingly, the Wino LSP scenario guarantees a mass splitting such that
the chargino could decay in any of the detector components.  This is
an automatic consequence of the Wino LSP scenario. With conventional
triggering, such Winos generally evade detection. For some range of
parameters, the splitting is such that Winos decay before the muon
chambers (although for long lifetimes those that do reach the muon
chamber will be important). Furthermore, the decay products are soft,
and will generally neither meet the calorimeter trigger threshold nor
provide an observable kink.  For short-lived tracks with sufficiently
hard decay products and for long-lived tracks, the current bound from
LEP II \cite{delphi} is about 90 GeV; otherwise it is 45--63 GeV.

Of course, if chargino events are accepted, the signal of a high $p_T$
track that disappears, leaving only a low momentum charged lepton or
pion, is spectacular, and could hardly escape off-line analysis.  The
essential difficulty then is the acceptance of chargino events into
the data sample.  In the following, we propose a number of solutions
to this difficulty and consider the prospects for probing the Wino LSP
scenario at the Fermilab detectors CDF II (Collider Detector at
Fermilab) and \Dzero\ (DZero) in the next Tevatron run.

We discuss several possible triggers. (I) For sufficiently long-lived
Winos, one can apply the usual search for heavy particles that trigger
in the muon chambers.  (II) For shorter-lived charginos which do not
reach the muon chamber, events in which a high $p_T$ jet accompanies
the Winos can be used by triggering on the jet and the associated
missing $E_T$. Distinguishing these events from background in the
off-line analysis will require identifying the Wino track itself.
Finally, as a supplement to these two triggers, we propose to search
for Winos too short-lived to reach the muon chamber by using the fact
that they leave stiff tracks in the tracking chamber in events that
are hadronically quiet.  This can be done by (III) triggering on
events with a single stiff track and no localized energy (in the form
of jets) in the calorimeter.  The addition of this trigger will extend
the Tevatron reach for the light Wino search and furthermore should
considerably enhance statistics.  A more conservative but less
powerful approach (III') would be to trigger instead on events
containing two stiff tracks with balancing $p_T$.  If $\Delta M$ is
signficantly above $m_\pi$, as for sequestered sector models,
only trigger II is useful, but in more general Wino LSP models all
three triggers can be important.

Trigger I is useful for detecting the processes
\begin{equation}
q \bar{q} \to \chargino^{\pm}_1 \neutralino_1,
\chargino^+_1 \chargino^-_1
\label{ww}
\end{equation}
when the Winos tracks have lengths of order meters or more.  Of
course, for the muon chamber trigger to be useful we must distinguish
Wino tracks from those produced by muons.  Fortunately, Winos tend to
have low velocities and associated high ionization energy loss rates
$dE/dx$ in the vertex detector and tracking chambers.  We will require
the Wino tracks to have $\beta\gamma < 0.85$, which corresponds to
$dE/dx$ approximately double minimally-ionizing~\cite{stauexp}.  In
Fig.~\ref{fig:ww}, we present the combined cross section for processes
(\ref{ww}), using the following technique.  Let $L$ be the minimum
radial distance a charged track must travel in order to be detected by
a given trigger (here, the distance to the muon chambers.)  We require
that each event have a charged track of length $L$ or greater, with
pseudorapidity $|\eta|<1.2$.  The cross section for such events
depends on $\Delta M$ through the combination $c\tau/L$.  We present
curves for several values of $c\tau/L$, with and without the cut on
$\beta\gamma$.  The figure shows that a cut on $\beta\gamma$ retains a
large signal, allowing Winos to be discovered in searches for massive
long-lived charged particles~\cite{stauth,stauexp}. The relative
sensitivity of this search depends, of course, on the chargino decay
length $c\tau$. For example, from Fig.~\ref{fig:ww}, we find that for
muon chambers with $L \approx 4.5$ m, assuming $2 \ \ifb$ integrated
luminosity and demanding 5 events for discovery, the mass reach for
Winos with $c\tau\ge 6$ m is at least 260 GeV.  Additional information
from time-of-flight may also be useful for distinguishing Winos from
muons.

\begin{figure}[tb]
\postscript{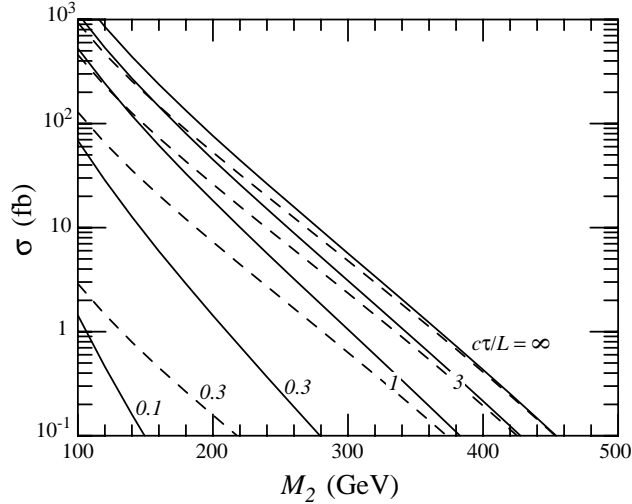}{0.51}
\caption{Cross sections (solid) at $\sqrt{s} = 2 \tev$ for Wino pair
production with at least one charged track traveling a radial length
$L$ with $|\eta| < 1.2$. The dependence on decay length $c\tau$ is
shown.  For the dashed contours, the charged track is also required to
have $\beta\gamma < 0.85$. See associated discussion of triggers I, III.}
\label{fig:ww}
\end{figure}

Next we consider trigger II, sensitive to the production of Winos plus
a jet.  Such topologies may be produced through the parton level
processes

\begin{equation}
q \bar{q} \to \chargino^{\pm}_1 \neutralino_1 g,
\chargino^+_1 \chargino^-_1 g \ \ {\rm and}\ \ q g \to
\chargino^{\pm}_1 \neutralino_1 q, \chargino^+_1 \chargino^-_1 q \ .
\label{qqjet}
\end{equation}
When the jet is hard, these events are characterized by large $\met$
resulting from a single high $p_T$ jet, and one or two charginos that
decay in the detector.  In our analysis, we require an event with
$\met>30$ GeV, and a jet with $p_T>30$ GeV and $|\eta|<1.2$.

For the signal to be distinguishable in the off-line analysis from
backgrounds, such as monojets resulting from $q \bar{q} \to gZ \to
g\nu \bar{\nu}$, the charginos, or their decay products, must be
visible.  The most obvious possibility is that the charginos leave
tracks in detector components before decaying.  We assume the off-line
analysis will require at least one isolated high $p_T$ track, with
$|\eta| < 2$, that travels a radial distance greater than some minimum
detection length $L$.  These tracks will not deposit much energy in
the calorimeters or (if short) hit the muon chambers, and should
therefore leave a spectacular, background-free signal.  (Note that in
events with long tracks that also hit the muon chambers, a cut on
$\beta\gamma$ will distinguish charginos from muons, as discussed
below.)

\begin{figure}[tb]
\postscript{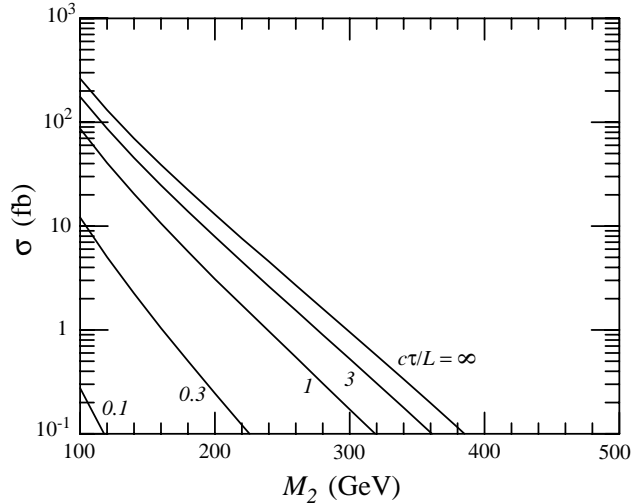}{0.51}
\caption{Cross sections at $\sqrt{s} = 2 \tev$ for associated
production of a Wino pair and a jet with $p_T > 30 \gev$ and $|\eta| <
1.2$.  At least one charged Wino is required to travel a radial length
$L$ with $|\eta| < 2$.  The dependence on decay length $c\tau$ is
shown. See associated discussion of trigger II.}
\label{fig:wwjet}
\end{figure}

In Fig.~\ref{fig:wwjet}, we plot cross sections, combining the four
relevant processes of (\ref{qqjet}) for various values of
$c\tau/L$. The cross sections are clearly strongly dependent on the
length $L$.  For both CDF II and \Dzero, a chargino traveling a radial
length $L=10$ cm or greater should be easily identified, as such
charginos will travel through essentially all layers of the silicon
vertex detector.  With the same discovery criterion as above, we find
a discovery reach of $M_2 \approx 140$, $210$, and $240 \gev$ for
decay lengths $c\tau = 3, 10, \text{ and } 30$ cm, respectively.

Winos with $c\tau<10$ cm decay predominantly through the pion mode.
If these pions can be identified, they could conceivably extend the
reach of this search for $\Delta M >m_\pi$. However, this requires
careful study outside the scope of this paper.

If the chargino track lengths are ${\cal O}(10\text{ cm})$ or longer,
trigger III could be applied to processes (\ref{ww}).  The rate for
chargino events accepted by such a trigger may be determined from the
solid curves in Fig.~\ref{fig:ww} for various $c\tau$.

As in the previous case, the cross sections depend strongly on the
required $L$.  For the CDF II (\Dzero) detector, tracking information
is available at the trigger level if $L\gtrsim$ 1 m (50 cm).  Once
such events are accepted, the lack of calorimeter activity makes them
striking; physics backgrounds are negligible, and the leading
backgrounds are expected to be instrumental.  (Long
tracks hitting the muon chamber will be discussed below.)  With the
same discovery criterion as above and $c\tau = 6$ m, both detectors
have a mass reach of roughly $320\gev$.  Furthermore, as can be seen
from Fig.~\ref{fig:ww}, for $c\tau\sim 6$ m the signal passing trigger
III ($c\tau/L\sim 10$) is several times larger than that passing
trigger I ($c\tau/L\sim 1$.)

Trigger III' accepts only the second process in (\ref{ww}), and
requires that both chargino tracks travel through a substantial portion
of the tracking chamber.  Though fewer signal events pass this
trigger, the ratio of signal to trigger background may be better than
for trigger III.  The utility of trigger III' is less than that of
trigger III, but is comparable to that of trigger I for $\Delta
M<m_\pi$.  If, contrary to our assumptions, the sleptons are not much
heavier than $M_2$, then the chargino lifetime would be smaller, and
the power of trigger I reduced, making trigger III' potentially more
important.

In our discussion of the discovery region for $\Delta M< m_\pi$, we
have neglected the fact that some fraction of the events passing
triggers II, III, and III' will contain Wino tracks that also pass
trigger I.  As before, these charginos must be distinguished from
muons using a $\beta\gamma$ cut.  However, most charginos are produced
slowly, so the impact of the $\beta\gamma$ cut is small, reducing the
discovery reach by at most 5--10 GeV.

In order to summarize the discovery reach, we show in
Fig.~\ref{fig:deltam} the 5 event discovery contours for triggers I,
II and III with $L$=4.5 m, 10 cm and 50 cm respectively.  In (a) we
have taken $M_1/M_2$ as suggested by the anomaly-mediated
supersymmetry breaking.  Since $\Delta M>m_\pi$, only trigger II plays
a role, but fortunately it can cover a large fraction of the parameter
space of the sequestered sector models.  In (b) we consider a more
general Wino LSP model in which the discovery reach is markedly
enhanced using triggers I and III.  In particular, triggers I and III,
which require small $\Delta M$ so that chargino tracks are
sufficiently long, are useful at large Wino masses where Wino
production is too rare for trigger II to find a signal.  Note that the
discovery reaches depend significantly on $M_1/M_2$, $\tan\beta$ and
${\rm sign}(\mu)$; these particular cases are for illustration only.

If candidate events are discovered, a number of important checks can
be made on the Wino LSP interpretation.  These include comparing the
number of events with one and two charged tracks, and determining the
fraction of events with anomalously large $dE/dx$ as mentioned above.
In addition, in order to distinguish this scenario from gauge-mediated
scenarios with long-lived sleptons, where macroscopic decay lengths
result not from degeneracy, but from highly suppressed
couplings~\cite{stauth,stauexp}, correlations between particle masses
and cross sections may be used.  Finally, as the signals discussed
above are essentially background-free, the discovery potential is
highly sensitive to integrated luminosity.  For example, if the total
luminosity is increased to $30\ifb$, the various Wino mass discovery
reaches estimated above increase by up to 100 GeV.  It is exciting
that Wino LSP searches will explore a large fraction of the parameter
space of the sequestered sector scenario~\cite{lr} even before the
LHC, giving the Tevatron the possibility of finding the first evidence
for extra space-time dimensions.

{\em Acknowledgments} --- It is a great pleasure to thank John Conway
and Darien Wood for many enlightening discussions.  SS thanks the
Institute for Advanced Study for hospitality.  This work was supported
in part by the Department of Energy under contracts
DE--FG02--90ER40542, DE--FG--02--91ER40671 and cooperative agreement
DF--FC02--94ER40818, the National Science Foundation under contract
NSF PHY--9513835, a Frank and Peggy Taplin Membership (JLF), a Marvin
L. Goldberger Membership (TM), and the W. M. Keck Foundation (MS).

\end{document}